\documentclass[aps,prd,reprint]{revtex4-2}
\usepackage{amsmath,amssymb}
\usepackage{graphicx}
\usepackage{hyperref}
\begin{document}
\title{Comment on ``Cosmological constraints on unimodular gravity models with diffusion'': thermodynamic inadmissibility of the $H_0$ tension resolution mechanism}\author{Mauricio~Cataldo}
\email{mcataldo@ubb.cl}
\affiliation{Departamento de F\'isica, Universidad del B\'io-B\'io, Casilla 5-C, Concepci\'on, Chile}
\date{\today} 
\begin{abstract}
We show that the diffusion-based models proposed in Refs.~\cite{Perez2021,Landau2022} within the framework of Unimodular Gravity (UG) to alleviate the $H_0$ tension are incompatible with the second law of thermodynamics. Starting from the Gibbs equation for a pressureless matter fluid, we derive a general thermodynamic admissibility condition for the $\Lambda$CDM$+$diffusion class in UG, demonstrating that the second law requires $\dot{Q} < 0$, independently of the specific form of the diffusion function $Q(t)$. This condition implies that energy must flow from the effective cosmological term $\Lambda_{\rm eff}$ into the matter sector, rather than in the opposite direction. We then establish a no-go theorem: no choice of $Q(t)$ can simultaneously satisfy the second law and generate the growing effective cosmological term $\dot{\Lambda}_{\rm eff} > 0$ required to alleviate the $H_0$ tension. We confirm this result explicitly for the models of Perez, Sudarsky, and Wilson-Ewing~\cite{Perez2021} and Landau et al.~\cite{Landau2022}, noting that the former explicitly identify $\dot{\Lambda}_{\rm eff} > 0$ as a necessary feature of their proposal without addressing its thermodynamic implications. The incompatibility is therefore structural and applies to the entire class of $\Lambda$CDM$+$diffusion models in UG with a pressureless matter component.
\end{abstract}

\maketitle

\section{Introduction}

The persistent tension between the value of the Hubble constant $H_0$ inferred from Cosmic Microwave Background (CMB) observations within the $\Lambda$CDM framework and the value measured directly from local distance indicators remains one of the most pressing open problems in modern cosmology~\cite{Verde2019,Riess2021,HuWang2023,Singh:2023jxu}. In this context, Unimodular Gravity~\cite{Bengochea:2023dep,Alvarez:2023utn,FernandezCristobal:2014jca} has attracted considerable attention as a natural framework to address this tension, since it allows for violations of the local conservation of the energy-momentum tensor without introducing any inconsistency in the field equations~\cite{Josset2017,Finkelstein:2000pg}.

The central idea shared by the proposals of Refs.~\cite{Perez2021,Landau2022} is that a diffusion process transfers energy from the matter sector into an effective dark energy component, modifying the late-time cosmological evolution and thereby reconciling the CMB-inferred and locally measured values of $H_0$. Both proposals are instances of what we shall call the \emph{$\Lambda$CDM$+$diffusion model in UG}: a spatially flat FLRW universe containing pressureless matter and an effective cosmological term $\Lambda_{\rm eff}(t) = \Lambda_0 + Q(t)$, where $\Lambda_0$ is a constant of integration and $Q(t)$ is a diffusion function encoding the non-conservation of the energy-momentum tensor. This model reduces exactly to standard $\Lambda$CDM of General Relativity when $Q = 0$. We stress that a defining feature of $\Lambda$CDM is that cold dark matter is pressureless, i.e., it has vanishing equation of state $w_{\rm cdm} = 0$~\cite{Peebles1982,Planck2018}; any model in which the dark matter component acquires an effective non-zero pressure falls outside this class by definition. The two proposals~\cite{Perez2021,Landau2022} differ only in the specific functional form assumed for $Q(t)$. We emphasize that the results of this comment, including the impossibility theorem proved in Sec.~IV, apply exclusively to this class of models and should not be interpreted as a general obstruction to the $H_0$ tension within the broader UG framework.

These proposals represent an important and creative line of research, and the results obtained are encouraging from the observational point of view. However, we note that none of these works verifies whether the adopted diffusion functions are compatible with the second law of thermodynamics. A thermodynamic analysis of UG with diffusion has recently been carried out in Ref.~\cite{Cruz2024}, where a general condition $T\,dS = -V\,dQ$ was derived and applied to the barotropic and continuous spontaneous localization diffusion models; however, that work does not discuss the $H_0$ tension and does not analyze the phenomenological models of Refs.~\cite{Perez2021,Landau2022}, which are precisely the ones proposed to resolve it. The present comment fills this gap by applying the thermodynamic admissibility condition to those specific models, showing that the very configurations required to alleviate the $H_0$ tension are thermodynamically inadmissible, and that this incompatibility is in fact structural: no choice of $Q(t)$ within the $\Lambda$CDM$+$diffusion model can simultaneously satisfy the second law and produce a growing $\Lambda_{\rm eff}$, as required by the $H_0$ resolution mechanism.

\section{The $\Lambda$CDM model with diffusion in UG}

We consider a spatially flat FLRW universe in UG, with pressureless matter of energy density $\rho_m$ and a diffusion function $Q(t)$. In the UG framework, the effective cosmological term is given by 
\begin{equation}
\Lambda_{\rm eff}(t) = \Lambda_0 + Q(t), \label{eq:Lambdaeff}
\end{equation}
where $\Lambda_0$ is a constant of integration and $Q(t)$ encodes the non-conservation of the energy-momentum tensor~\cite{Josset2017}. The background equations are
\begin{align}
3H^2 &= 8\pi G\,\rho_m + \Lambda_0 + Q, \label{eq:friedmann}\\
2\dot{H} + 3H^2 &= -8\pi G\,p_m + \Lambda_0 + Q, \label{eq:raychaudhuri}\\
\dot{\rho}_m + 3H\rho_m &= -\dot{Q}, \label{eq:continuity}
\end{align}
where $H = \dot{a}/a$ is the Hubble rate and $p_m = 0$ because $\rho_m$ describes cold dark matter (CDM), which is by definition a pressureless fluid with vanishing equation of state $w_{\rm cdm} = 0$~\cite{Peebles1982,Planck2018}. Equation~\eqref{eq:continuity} thus represents the continuity equation for a pressureless CDM fluid in the presence of diffusion. In this equation, the direction of the energy transfer is determined by the sign of $\dot{Q}$: $\dot{Q} > 0$ implies $\dot{\Lambda}_{\rm eff} > 0$, corresponding to an energy flow from the matter sector into $\Lambda_{\rm eff}$; conversely, $\dot{Q} < 0$ implies an energy transfer from $\Lambda_{\rm eff}$ into the matter sector.

We note that the decomposition~\eqref{eq:Lambdaeff}, often used in the literature, is flexible enough to accommodate a variety of physical situations while keeping both $\Lambda_0 > 0$ and $\Lambda_{\rm eff}(t) > 0$ at all times: since $\Lambda_0$ is a fixed positive integration constant, the diffusion function $Q(t)$ may start from a negative value and grow with time, provided $Q(t) > -\Lambda_0$ throughout the evolution. This sign structure will be relevant when comparing the models of Refs.~\cite{Perez2021} and~\cite{Landau2022} in Sec.~III.

We emphasize that Eq.~\eqref{eq:continuity} constitutes the defining structural feature of the $\Lambda$CDM$+$diffusion class: all models analyzed in Sec.~III that alleviate the $H_0$ tension while violating the second law of thermodynamics do so precisely because their matter density obeys Eq.~\eqref{eq:continuity} with $p_m = 0$, from which, as we will show, the thermodynamic condition $T\dot{S} = -a^3\dot{Q}$ follows directly. Note that this does not imply that $\rho_m$ necessarily evolves as $a^{-3}$: for instance, in the Anomalous Decay model the solution is $\rho_m \propto a^{-(3+\gamma)}$ with $\gamma > 0$. What these models share is not the $a^{-3}$ scaling but the form of Eq.~\eqref{eq:continuity}, from which the thermodynamic condition follows independently of the specific solution for $\rho_m(t)$ or the functional form of $Q(t)$. 

When $Q = 0$, Eqs.~\eqref{eq:friedmann}--\eqref{eq:continuity} reduce exactly to the standard $\Lambda$CDM model of General Relativity, with $\Lambda_0$ playing the role of the cosmological constant. This is the key property that motivates the use of UG as a framework to modify $\Lambda$CDM: the diffusion function $Q(t)$ introduces departures from the standard evolution while preserving the $\Lambda$CDM limit.

The physical mechanism proposed in Refs.~\cite{Perez2021,Landau2022} consists of an effective energy flux from the matter sector into the dark energy sector, which leads to an increasing $Q(t)$, i.e., $\dot{Q} > 0$. This growth in $Q$ alters the late-time expansion rate and constitutes the mechanism proposed to alleviate the $H_0$ tension.

\subsection{Thermodynamic constraint on $Q(t)$}

We now derive the condition that $Q(t)$ must satisfy in order for the model to be thermodynamically admissible. As shown in Ref.~\cite{Cruz2024}, this condition is independent of the specific functional form of $Q(t)$. For completeness we reproduce the derivation here before applying it to the models of Refs.~\cite{Perez2021,Landau2022}.

For a pressureless fluid in an expanding FLRW universe, the Gibbs equation reads
\begin{equation}
T\,dS = dU + p_m\,dV = d(\rho_m V),
\end{equation}
where $V = a^3$ is the physical volume of a unit comoving region, $p_m = 0$, and $U = \rho_m V = \rho_m a^3$ is the total energy contained in that region. Taking the time derivative,
\begin{equation}
T\dot{S} = a^3\left(\dot{\rho}_m + 3H\rho_m\right).
\end{equation}
Substituting the continuity equation~\eqref{eq:continuity},
\begin{equation}
T\dot{S} = -a^3\,\dot{Q}.
\label{eq:entropy}
\end{equation}
Since $T > 0$ and $a^3 > 0$, the second law of thermodynamics $\dot{S} > 0$ imposes the following necessary condition on the diffusion function:
\begin{equation}
\dot{Q} < 0.
\label{eq:2ndlaw}
\end{equation}
Equivalently, in terms of redshift $z$, defined via $a = (1+z)^{-1}$ so that $\dot{Q} = -H(1+z)\,dQ/dz$, the condition reads
\begin{equation}
\frac{dQ}{dz} > 0,
\label{eq:2ndlaw_z}
\end{equation}
that is, $Q$ must be a decreasing function of cosmic time, or equivalently an increasing function of redshift. This result holds for any choice of $Q(t)$ and constitutes the fundamental thermodynamic admissibility condition for the $\Lambda$CDM$+$diffusion model in UG.

The thermodynamic admissibility condition~\eqref{eq:2ndlaw_z} is fully consistent with the results of Refs.~\cite{Cruz2024,CruzLepePalma2026}. Cruz, Lepe and Palma~\cite{CruzLepePalma2026}, working with a single barotropic fluid and the ansatz $Q(z) = Q_0(1+z)^\beta$, showed that the second law imposes $\beta Q_0 > 0$, which translates directly into $dQ/dz > 0$, in complete agreement with condition~\eqref{eq:2ndlaw_z}; the thermodynamically inadmissible regime corresponds to $\beta Q_0 < 0$, i.e., $dQ/dz < 0$.

\section{Thermodynamic analysis of the proposed models}

We now show that the two proposals of Refs.~\cite{Perez2021,Landau2022} are both instances of the general $\Lambda$CDM$+$diffusion model described in Sec.~II, each corresponding to a specific choice of $Q(t)$. We then verify condition~\eqref{eq:2ndlaw} for each case.

\subsection{Models of Perez, Sudarsky and Wilson-Ewing}

Reference~\cite{Perez2021} proposes two phenomenological models, both
instances of the general $\Lambda$CDM$+$diffusion framework of Sec.~II
with specific choices of $Q(z)$.

\textit{Model A: Sudden transfer.} The matter density undergoes a sudden
decrease at redshift $z^*$,
\begin{eqnarray}
\rho_m(z) &=& \rho_m^0(1+z)^3\left[\theta_+(z-z^*) +
\alpha\,\theta_-(z-z^*)\right],
\end{eqnarray}
where $0< \alpha < 1$. Substituting into Eq.~\eqref{eq:continuity} and using
$Q = \Lambda_{\rm eff} - \Lambda_0$, the diffusion function reads
\begin{equation}
Q(z) = \hat{\Delta}\,H_0^2\,\theta_-(z-z^*), \qquad
\hat{\Delta} = 3(1-\alpha)(1+z^*)^3\bar{\Omega}_m^0 > 0,
\label{eq:QA}
\end{equation}
where $\theta_-(z-z^*) = 1$ for $z < z^*$ and zero otherwise, and $\hat{\Delta} > 0$ follows from $\alpha < 1$. In this model $Q$ starts at zero in the distant past and jumps to a positive value $\hat{\Delta}H_0^2$ at $z^*$, remaining positive until today;
consequently $\Lambda_{\rm eff} = \Lambda_0 + \hat{\Delta}H_0^2 >
\Lambda_0 > 0$ for $z < z^*$, as anticipated in Sec.~II. As $z$
decreases through $z^*$, $\Lambda_{\rm eff}$ grows from $\Lambda_0$ to
$\Lambda_0 + \hat{\Delta}H_0^2$, i.e., $\dot{\Lambda}_{\rm eff} > 0$,
in direct violation of condition~\eqref{eq:2ndlaw_z}. As
Eq.~\eqref{eq:entropy} shows, this implies $T\dot{S} < 0$, a
consequence that Ref.~\cite{Perez2021} does not discuss.

\textit{Model B: Anomalous decay.} The diffusion function for
$z \leq z^*$ is
\begin{equation}
Q(z) = -\frac{3\gamma\bar{\Omega}_m^0}{\gamma+3}
\left[\left(\frac{1+z}{1+z^*}\right)^\gamma(1+z)^3 -
(1+z^*)^3\right]H_0^2,
\label{eq:QB}
\end{equation}
with $Q = 0$ for $z > z^*$. A direct computation gives
\begin{equation}
\frac{dQ}{dz} = -3\gamma\bar{\Omega}_m^0\,
\frac{(1+z)^{\gamma+2}}{(1+z^*)^\gamma}\,H_0^2 < 0
\end{equation}
for all $\gamma > 0$, i.e., $\dot{\Lambda}_{\rm eff} > 0$ throughout
$z \leq z^*$. As in Model~A, Eq.~\eqref{eq:entropy} then gives
$T\dot{S} < 0$, violating condition~\eqref{eq:2ndlaw_z}.

We note that this thermodynamic condition is in fact consistent with the analysis of Ref.~\cite{Perez2021} itself: in Sec.~V of that work, when analyzing the jerk parameter $j_0 = 1 + \dot{\Lambda}(0)/2H_0^3$, Perez et al.\ argue that the energy transfer from matter to dark energy implies $\dot{\Lambda} > 0$ and hence $j_0 > 1$, thereby explicitly identifying $\dot{\Lambda}_{\rm eff} > 0$ as a necessary feature of their proposal. However, the thermodynamic consequence of this condition, namely that $\dot{\Lambda}_{\rm eff} > 0$ implies $T\dot{S} < 0$ via Eq.~\eqref{eq:entropy}, in direct violation of the second law, is not discussed in Ref.~\cite{Perez2021}. Therefore, in both models, alleviating the $H_0$ tension requires $\dot{Q} > 0$, which violates condition~\eqref{eq:2ndlaw}.

\subsection{Model of Landau et al.}

Reference~\cite{Landau2022} extends the preliminary analysis of Ref.~\cite{Perez2021} by performing a full Boltzmann-code treatment that includes both the background evolution and the growth of perturbations, comparing the model predictions with a comprehensive dataset comprising CMB anisotropies and polarization from Planck 2018, baryon acoustic oscillations, Type Ia supernovae (Pantheon compilation), and cosmic chronometers. The physical motivation remains the same: an effective energy flux from the matter sector (baryons and cold dark matter) into $\Lambda_{\rm eff}$, attributed to rotational friction of black holes produced by space-time granularity at the Planck scale.

The model is parametrized by three quantities: the amplitude $\Delta\rho_\Lambda \equiv \Delta\rho\,h^2/\rho_{\rm crit} = 8\pi G\,\Delta\rho/(3\times 100^2)$, where $h = H_0/(100\,{\rm km\,s^{-1}\,Mpc^{-1}})$, characterizing the magnitude of the energy transfer, the scale factor $a^*$ at which the transfer occurs, and the duration $\delta$ of the transfer window. The effective cosmological term is postulated to evolve as $\rho_\Lambda(a) = \rho_\Lambda^0 + \Delta\rho\,[f(a)-1]$, where
\begin{equation}
f(a) = \begin{cases}
0 & a < a^* - \delta/2, \\
\dfrac{a - a^* + \delta/2}{\delta} & a^* - \delta/2 \leq a \leq a^* + \delta/2, \\
1 & a > a^* + \delta/2.
\end{cases}
\end{equation}
Translating into the language of Sec.~II, the corresponding diffusion function reads
\begin{equation}
Q(a) = \Delta\rho\left[f(a) - 1\right],
\end{equation}
so that $Q = -\Delta\rho < 0$ for $a < a^* - \delta/2$, $Q$ increases linearly from $-\Delta\rho$ to $0$ in the transition window, and $Q = 0$ for $a > a^* + \delta/2$. This sign structure is precisely the flexible scenario described in Sec.~II: $Q$ starts from a negative value in the past and grows toward zero today, while $\Lambda_{\rm eff} = \Lambda_0 + Q$ remains positive throughout provided $\Delta\rho < \Lambda_0$, and the diffusion is completely switched off in the present epoch. For $\Delta\rho > 0$, which is the configuration that alleviates the $H_0$ tension, $Q$ is an increasing function of $a$, i.e., a decreasing function of redshift $z$. Therefore $dQ/dz < 0$, in direct violation of the thermodynamic admissibility condition~\eqref{eq:2ndlaw_z}.

The statistical analysis of Ref.~\cite{Landau2022} explores $\Delta\rho_\Lambda \in [-0.2, 0.2]$, but only the region $\Delta\rho_\Lambda > 0$ is thermodynamically inadmissible and coincides with the configurations that ease the $H_0$ tension. The main analysis, with all unimodular parameters free to vary, yields $\Delta\rho_\Lambda = 0.0010 \pm 0.0040$ — consistent with zero — and $H_0 = 68.51 \pm 0.57\,{\rm km\,s^{-1}\,Mpc^{-1}}$, providing only a marginal improvement over the $\Lambda$CDM value of $H_0 = 68.02 \pm 0.40\,{\rm km\,s^{-1}\,Mpc^{-1}}$ obtained with the same dataset. A speculative analysis with the unimodular parameters fixed at $\Delta\rho_\Lambda = 0.09$, $a^* = 0.4$, $\delta = 0.02$ yields $H_0 = 71.18 \pm 0.42\,{\rm km\,s^{-1}\,Mpc^{-1}}$, in agreement within $1.9\sigma$ with the local measurement of Ref.~\cite{Riess2021}. However, this improved agreement is achieved precisely by fixing $\Delta\rho_\Lambda > 0$, which corresponds to the thermodynamically inadmissible regime $\dot{Q} > 0$ identified above, thereby violating condition~\eqref{eq:2ndlaw}.

\section{Discussion and an impossibility theorem}

The analysis presented in the preceding section shows that both proposals in Refs.~\cite{Perez2021,Landau2022} require $\dot{Q} > 0$, i.e., an increasing $\Lambda_{\rm eff}(t)$, in order to alleviate the $H_0$ tension. This requirement is in direct contradiction with the thermodynamic condition~\eqref{eq:2ndlaw} established in Sec.~II. The common mechanism in both proposals is an effective energy flux from the matter sector into $\Lambda_{\rm eff}$, which increases the late-time expansion rate and reconciles the CMB-inferred and locally measured values of $H_0$. This mechanism is precisely characterized by $\dot{\Lambda}_{\rm eff} > 0$, or equivalently $\dot{Q} > 0$. The conflict between thermodynamic admissibility and this specific mechanism of $H_0$ resolution is not merely an accident of the specific forms of $Q$ chosen in those works. It is, in fact, a structural incompatibility that applies to the entire class of $\Lambda$CDM$+$diffusion models in UG with a pressureless matter fluid. We formalize this as the following result.

\medskip
\noindent\textbf{Theorem.} \textit{In the $\Lambda$CDM$+$diffusion framework of UG described by Eqs.~\eqref{eq:friedmann}--\eqref{eq:continuity}, with a pressureless matter fluid ($p_m = 0$), there exists no diffusion function $Q(t)$ that simultaneously satisfies the second law of thermodynamics and produces a growing effective cosmological term, $\dot{\Lambda}_{\rm eff} = \dot{Q} > 0$. Consequently, the specific mechanism of $H_0$ tension resolution adopted in Refs.~\cite{Perez2021,Landau2022}, namely, an energy flux from matter into $\Lambda_{\rm eff}$, is thermodynamically inadmissible within this framework, regardless of the functional form chosen for $Q(t)$.}

\medskip
\noindent\textit{Proof.} From Eq.~\eqref{eq:entropy}, $T\dot{S} = -a^3\dot{Q}$. Since $T > 0$ and $a^3 > 0$, the second law $\dot{S} > 0$ requires $\dot{Q} < 0$, i.e., $Q$ must decrease with cosmic time, which implies $\dot{\Lambda}_{\rm eff} = \dot{Q} < 0$. The condition $\dot{Q} > 0$ required to grow $\Lambda_{\rm eff}$ is therefore mutually exclusive with the second law, for any choice of $Q(t)$. $\square$

\medskip

The physical content of this theorem is transparent. The second law demands that the matter fluid gains entropy, which requires energy to flow \emph{into} the matter sector, i.e., $\dot{Q} < 0$. But the $H_0$ resolution mechanism of Refs.~\cite{Perez2021,Landau2022} requires energy to flow \emph{out of} the matter sector and into $\Lambda_{\rm eff}$, i.e., $\dot{Q} > 0$. These two directions of energy flow are opposite and cannot be reconciled within this framework. 

We emphasize that this result does not preclude alternative mechanisms for addressing the $H_0$ tension within UG that do not rely on a growing $\Lambda_{\rm eff}$. Such mechanisms, operating through different physical channels, may remain thermodynamically admissible and therefore constitute viable directions for future investigation. The present analysis establishes thermodynamic consistency as a necessary constraint that any future diffusion-based model in UG must satisfy alongside observational viability.

\begin{acknowledgments}
M.C. acknowledges support from the Direcci\'on de Investigaci\'on y Creaci\'on Art\'istica at the Universidad del B\'io-B\'io through grant GI2310339.
\end{acknowledgments}


\begin{thebibliography}{99}

\bibitem{Verde2019}
L.~Verde, T.~Treu, and A.~G.~Riess,
\textit{Tensions between the early and late Universe},
Nat.\ Astron.\ \textbf{3}, 891 (2019),
arXiv:1907.10625.

\bibitem{Riess2021}
A.~G.~Riess et al.,
\textit{A Comprehensive Measurement of the Local Value of the Hubble Constant with 1 km/s/Mpc Uncertainty from the Hubble Space Telescope and the SH0ES Team},
Astrophys.\ J.\ Lett.\ \textbf{934}, L7 (2022),
arXiv:2112.04510.

\bibitem{HuWang2023}
J.-P.~Hu and F.-Y.~Wang,
\textit{Hubble Tension: The Evidence of New Physics},
Universe \textbf{9}, 94 (2023),
arXiv:2302.05709.

\bibitem{Singh:2023jxu}
N.~K.~Singh and G.~Kashyap,
\textit{Unimodular Theory of Gravity in Light of the Latest Cosmological Data},
Universe \textbf{9}, 469 (2023),
arXiv:2306.17754.

\bibitem{Bengochea:2023dep}
G.~R.~Bengochea, G.~Leon, A.~Perez, and D.~Sudarsky,
\textit{A clarification on prevailing misconceptions in unimodular gravity},
JCAP \textbf{11}, 011 (2023),
arXiv:2308.07360.

\bibitem{Alvarez:2023utn}
E.~Alvarez and E.~Velasco-Aja,
\textit{A Primer on Unimodular Gravity},
arXiv:2301.07641 (2023).

\bibitem{FernandezCristobal:2014jca}
J.~M.~Fern{\'a}ndez Crist{\'o}bal,
\textit{Unimodular theory: A little pedagogical vision},
Annals Phys.\ \textbf{350}, 441 (2014).

\bibitem{Josset2017}
T.~Josset, A.~Perez, and D.~Sudarsky,
\textit{Dark Energy from Violation of Energy Conservation},
Phys.\ Rev.\ Lett.\ \textbf{118}, 021102 (2017),
arXiv:1604.04183.

\bibitem{Finkelstein:2000pg}
D.~R.~Finkelstein, A.~A.~Galiautdinov, and J.~E.~Baugh,
\textit{Unimodular relativity and cosmological constant},
J.\ Math.\ Phys.\ \textbf{42}, 340 (2001),
arXiv:gr-qc/0009099.

\bibitem{Perez2021}
A.~Perez, D.~Sudarsky, and E.~Wilson-Ewing,
\textit{Resolving the $H_0$ tension with diffusion},
Gen.\ Rel.\ Grav.\ \textbf{53}, 7 (2021),
arXiv:2001.07536.

\bibitem{Landau2022}
S.~J.~Landau, M.~Benetti, A.~Perez, and D.~Sudarsky,
\textit{Cosmological constraints on unimodular gravity models with diffusion},
Phys.\ Rev.\ D \textbf{108}, 043524 (2023),
arXiv:2211.07424.

\bibitem{Peebles1982}
P.~J.~E.~Peebles,
\textit{Large-scale background temperature and mass fluctuations due to scale-invariant primeval perturbations},
Astrophys.\ J.\ \textbf{263}, L1 (1982).

\bibitem{Planck2018}
Planck Collaboration, N.~Aghanim et al.,
\textit{Planck 2018 results. VI. Cosmological parameters},
Astron.\ Astrophys.\ \textbf{641}, A6 (2020),
arXiv:1807.06209.

\bibitem{Cruz2024}
M.~Cruz, N.~Cruz, and S.~Lepe,
\textit{Exploring thermodynamics inconsistencies in unimodular gravity: a comparative study of two energy diffusion functions},
Eur.\ Phys.\ J.\ C \textbf{84}, 1186 (2024),
arXiv:2407.15207.

\bibitem{CruzLepePalma2026}
N.~Cruz, S.~Lepe, G.~Palma, and M.~Cruz,
\textit{Thermodynamic constraints and future singularities in Unimodular Gravity driven by phantom and non-phantom fluids},
arXiv:2603.22749 (2026).
\end{thebibliography}
\end{document}